\documentclass[preprint,pra,aps,showpacs]{revtex4}
\begin{document}

\title{An equilibrium state with macroscopic correlations}

\author{Giada Basile}
\affiliation{Dipartimento di Fisica, Universit\`a ``La Sapienza''\\
Piazzale Aldo moro 2, 00185 Roma, Italy}
\author{Giovanni Jona-Lasinio}
\affiliation{Dipartimento di Fisica and INFN, 
Universit\`a ``La Sapienza''\\
Piazzale Aldo Moro 2, 00185 Roma, Italy}

\begin{abstract}
In this paper we show that the equilibrium macroscopic entropy of a 
generic non reversible {\sl Kawasaki+Glauber} dynamics
is a non local functional of the density. This implies that
equilibrium correlations extend to macroscopic distances.
\bigskip
\end{abstract}

\maketitle


\section{Introduction}
In the last two years considerable progress has been made in understanding
stationary non equilibrium states (SNS). An important result, 
first established by Derrida, Lebowitz and Speer \cite{DLS1, DLS} 
and then  rederived in a simpler way from general principles 
in \cite{BDGJL}, 
is the non local character of the entropy in a boundary driven 
(i.e. interacting with reservoirs having different chemical
potentials) simple exclusion system. The question naturally arises whether
this is a general property of SNS. A different model, the boundary driven
zero range, does not exhibit this property \cite{BDGJL1, BDGJL} but one may suspect that this is related to the special character 
of the interaction.
All these models are conservative in the sense that the number of particles is conserved.

In this note we present an example of an equilibrium state in which the
entropy is a non local functional of the thermodynamic variables showing
in this way that this is not an exclusive property of SNS. 
By an equilibrium state we mean a stationary state of an isolated system
or of a system in contact with reservoirs characterized by the same
chemical potential. Our conclusion is that non locality or 
macroscopic correlations is more likely to be a generic feature of non 
reversible dynamics. Indeed, in the model considered, the generator 
of the dynamics
is not self-adjoint with respect to the equilibrium measure. This means
that detailed balance does not hold for such an equilibrium ensemble.

The system considered evolves according to the so called 
{\sl Kawasaki+Glauber} dynamics with generic transition rates 
for the {\sl Glauber} process and will be described  in the next 
section.  
Our analysis is based
on a general Hamilton-Jacobi equation satisfied by the entropy 
in equilibrium or non equilibrium states and established in \cite{BDGJL1, BDGJL}. The H-J equation for our model has a very complicated structure 
but for values of the
thermodynamic variables near the stationary values can be solved
by an iterative scheme of calculation. The entropy is non local already
in the lowest approximation, i.e. at the level of gaussian fluctuations.
One should mention that there are special 
choices of the transition probabilities which make the entropy of the 
model local in spite of a microscopic non reversible dynamics \cite{GJLV},
however we are interested in properties with some degree of
stability.
A more detailed analysis of our example and of the non locality  
problem will be presented elsewhere. 

\section{The model}
In this section we follow closely \cite{GJLV}. 
The model consists of particles on a lattice evolving according to two
basic dynamical processes:

\begin{itemize}
\item
i) a particle can move to a neighbouring site if this is empty
\item
ii) a particle can disappear or be created in a site, the
rate depending on the nearby configuration. 
\end{itemize}
The first process is conservative while the second is not.

Mathematically we consider a family of Markov processes whose state space
is $X_N=\{0,1\}^{Z_N}$,
where $N$ is an integer and $Z_N$ denotes the set of integers modulo $N$.
We shall denote with $\eta$ a point in the state space, that is a 
configuration of the system. This is therefore given by a function 
$\eta(i)$ defined on each site and taking the values $0$ or $1$.
For each $N$ the dynamics  is defined by the action of
the Markov generator $L_N$  on functions $f(\eta)$ 
\begin{equation}
L_Nf(\eta)=\frac{N^2}{2}\sum_{i\in Z_N}(f(\eta^{i,i+1})-f(\eta))+
\sum_{i\in Z_N}c(i,\eta)(f(\eta^i)-f(\eta))
\label{eqn1}
\end{equation}
where the addition in $Z_N$ means addition modulo $N$ 
\begin{equation}
\eta^{i,k}(j)= \left\{
\begin{array}{ccl}
\eta(j) & & j\neq i,k \\
\eta(k) & & j=i \\
\eta(i) & & j=k 
\end{array}
\right.
\label{eqn2}
\end{equation}
\begin{equation}
\eta^i(j)=\left\{
\begin{array}{ccl}
\eta(j) & & j\neq i \\
1-\eta(i) & & j=i
\end{array}
\right.
\label{eqn3}
\end{equation}
The rates $c(i,\eta)$ depend on the values of $\eta(j)$ with $j$ 
within a fixed distance $R$ from the site $i$. They are translation 
invariant, that is there exists a function $c(\eta)$ such
that  $c(i,\eta)=c(\tau_i\eta)$ where  
$(\tau_k\eta)(j)=\eta(j-k)$.
Let us consider now the unit interval $[0,1)$ with periodic condition 
at the  boundary and a function $\gamma$ defined on $[0,1)$ and taking 
values in $[0,1]$.
Let $\nu_{\gamma}^N$ the probability measure on the state space of 
the system obtained by assigning a Bernoulli distribution to each site, 
taking the product over all sites and defined by
\begin{equation}
\nu_{\gamma}^N\{\eta(k)=1\}=\gamma (\frac{k}{N})
\label{eqn4}
\end{equation}

The main object of our study is the empirical density $\mu_t^N$:
\begin{equation}
\mu_t^N(x)=\frac{1}{N}\sum_{k\in Z_N}\eta_{N^2t}(k)\delta(x-\frac{k}{N})
\label{eqn5}
\end{equation}
Let us denote by $Q_{\gamma}^N$ the distribution law 
of the trajectories $\mu_t^N(x)$  when the initial measure 
is the product measure $\nu^N_\gamma$.
It is possible to show that
$Q_{\gamma}^N$ converges weakly as $N$ goes to infinity 
to the  measure concentrated on
the path $\rho(t,x)$ that is the unique solution of
\begin{equation}
\left\{
\begin{array}{ccl}
\partial_t \rho &=& \frac{1}{2}\partial_x^2\rho +B(\rho)-D(\rho) \\
\rho(0,\cdot) &=& \gamma(\cdot)
\end{array}
\right.
\label{eqn6}
\end{equation}
with
\begin{equation}
B(\rho)=E_{\nu_{\rho}}(c(\eta)(1-\eta(0)))
\label{eqn11}
\end{equation}
\begin{equation}
D(\rho)=E_{\nu_{\rho}}(c(\eta)\eta(0))
\label{eqn12}
\end{equation}
where $\nu_{\rho}$ is the Bernoulli product distribution with 
$\gamma(x)\equiv \rho$.
Typically $B(\rho)$ and $D(\rho)$ are polynomials in the variable $\rho$.

The equilibrium state corresponds to a density $\bar \rho$ which is
the solution of the equation $B(\rho)=D(\rho)$ that gives an absolute
minimum of the potential $V(\rho)=\int^{\rho}[D(\rho') - B(\rho')]d\rho'$.

The above result is a law of large numbers that shows that the empirical
density in the limit of large $N$ behaves deterministically according to
 equation (\ref{eqn6}). 
We can now ask what is the probability that $\mu^N_t$ follows a 
trajectory different from a solution of (\ref{eqn6}) when $N$ is large 
but not infinite. This probability is exponentially small in $N$ and 
can be estimated using the methods of the theory of large deviations 
introduced for the systems of interest in \cite{KOV} and developed in 
\cite{L, GJLV}. The main idea consists in introducing a modified
process for which the trajectory of interest (fluctuation) is  
a solution of the corresponding hydrodynamic equation, and then 
comparing the two evolutions. 

For this purpose we consider the Markov process defined by the generator   
\begin{equation}
\left.\begin{array}{ccl}
L_{N,t}^Hf(\eta)&=&\frac{N^2}{2}\sum_{|i-j|=1}\eta(i)(1-\eta(j))
e^{H(t,\frac{j}{N})-H(t,\frac{i}{N})}[f(\eta^{i,j})-f(\eta)] \\
&+& \sum_i c(i,\eta)[(1-\eta(i))e^{H(t,\frac{i}{N})}+\eta(i)
e^{-H(t,\frac{i}{N})}][f(\eta^i)-f(\eta)]
\end{array}\right.
\label{eqn9}
\end{equation}
with $c$, $\eta^{k,j}$, $\eta^i$ as previously defined and
$H(t,x)$ can be interpreted as an external potential.

The deterministic equation satisfied by the empirical density is now
\begin{equation} 
\left\{
\begin{array}{ccl}
\partial_t \rho &=& \frac{1}{2}\partial_x^2\rho-\partial_x(\rho(1-\rho)
\partial_xH)+B(\rho)e^H-D(\rho)e^{-H} \\
\rho(0,\cdot) &=& \gamma(\cdot)
\end{array}
\right.
\label{eqn10}
\end{equation}

Given a function $\rho(x,t)$ twice differentiable with respect to $x$ 
and once with respect to $t$ this equation determines uniquely 
the field $H$. 
The probability that $\mu^N_t$ in the unperturbed system 
follows a trajectory different from a solution of (\ref{eqn6}) 
can now be expressed in terms of  
$H$ and the polynomials $B$ and $D$. We introduce the large 
deviation functional 
\begin{equation}
\left.\begin{array}{ccl}
I(\rho_t) &=& \frac{1}{2}\int_0^{t_0}\int_0^1dtdx\rho_t(1-\rho_t)
(\partial_xH_t)^2 \\
&+& \int_0^{t_0}\int_0^1dtdxB(\rho_t)(1-e^{H_t}+H_te^{H_t}) \\
&+& \int_0^{t_0}\int_0^1dtdxD(\rho_t)(1-e^{-H_t}-H_te^{-H_t})
\end{array}\right.
\label{eqn14}
\end{equation}
Let $\rho_t$ be a  trajectory in the interval of time $[0,t_0]$
with initial profile $\gamma$; assume that the initial 
configuration of the process $\eta_0$ is fixed and   
such that $\mu^N_0(x)\rightarrow \gamma(x)$ as $N\rightarrow \infty$. 
The large fluctuation estimate asserts that 
\begin{equation}
P_{\eta_0}^N(\mu_t^N \sim \rho_t, t\in [0,t_0])\simeq e^{-NI(\rho_t)} 
\label{lf}
\end{equation}
where $P_{\eta_0}^N$ is the probability distribution of the 
unperturbed process $\eta_t$ starting in $\eta_0$.
The sign $\simeq$ has to be interpreted as asymptotic equality of the 
logarithms.
Equation (\ref{lf}) is a dynamical generalization
of Einstein formula for thermodynamic fluctuations.
From this equation one sees that  
the trajectory that creates a certain profile $\rho(x)$ 
with highest probability is the one 
that minimizes $I(\rho_t)$ in the set 
$G$ of all trajectories that connect the equilibrium state $\bar \rho$
to $\rho(x)$. 

Let us introduce the quantity
\begin{equation}
S(\rho)=\inf_{\rho_t \in G}I(\rho_t)
\label{inf}
\end{equation}
that we shall
call the entropy associated to the profile $\rho$. 
One can show, from the large deviation estimate (\ref{lf}), that
in equilibrium 
\begin{equation}
P_{eq}^N(\mu^N \sim \rho)\simeq e^{-NS(\rho)} 
\label{olf}
\end{equation} 
In this way we recover dynamically the usual Einstein formula.
Note that the sign
of the entropy is the opposite of that used in the physical literature. 

\section{The Hamilton-Jacobi equation and its consequences}
The functional $I(\rho)$ has the form of an action integral associated 
to the Lagrangian 
\begin{equation}
\left.\begin{array}{ccl}
{\cal L} (\rho, \dot \rho) &=& \frac{1}{2}\int_0^1dx\rho_t(1-\rho_t)
(\partial_xH_t)^2 \\
&+& \int_0^1dxB(\rho_t)(1-e^{H_t}+H_te^{H_t}) \\
&+& \int_0^1dxD(\rho_t)(1-e^{-H_t}-H_te^{-H_t})
\end{array}\right.
\label{LAG}
\end{equation}
where $H(\rho, \dot \rho)$ is determined by (\ref{eqn10}). 
By Legendre transform we can define the Hamiltonian
\begin{equation}
{\cal H}(\rho, H)=\int_0^1 dx \big(\frac {1}{2} H\partial_x^2 \rho  + \frac {1}{2}(\partial_x H)^2 \rho(1-\rho)
- B(\rho)(1-\exp{H}) - D(\rho)(1-\exp{-H}) \big)
 \label{HA}
\end{equation}
The entropy $S(\rho)$ then satisfies the Hamilton-Jacobi equation
\begin{equation}
{\cal H}(\rho, \frac {\delta S}{\delta \rho})=0
\label{HJ}
\end{equation}
This is a very complicated functional derivative equation which however
can be solved by successive approximations using as an expansion parameter
$\rho - \bar \rho$ where $\bar \rho$ is a solution of $B(\rho)=D(\rho)$
that is a stationary solution of hydrodynamics. For $\rho=\bar \rho$
we have $\frac {\delta S}{\delta \rho} = 0$. We are looking for
an  approximate solution of (\ref{HJ}) of the form
\begin{equation}
S(\rho) = \frac {1}{2} \int_0^1 dx \int_0^1 dy (\rho(x) - \bar \rho)k(x,y)(\rho(y) - \bar \rho) + o(\rho - \bar \rho)^2
\end{equation}

 The kernel $k(x,y)$ is the inverse of the density correlation
function. It is more convenient, as in \cite{BDGJL},
to work directly with the
correlation function using the Legendre transform of the entropy
$G(h)$, which is the pressure corresponding to the chemical
potential profile $h$
\begin{equation}
 G(h)=\sup_\rho\{\langle h,\rho\rangle-S(\rho)\}
\end{equation}
G(h) satisfies the dual Hamilton-Jacobi equation
\begin{equation}\label{H_J,G}
\mathcal{H}\left(\frac{\delta G}{\delta h},h\right)=0
\end{equation}
We are looking for an approximate solution of (\ref{H_J,G}) of the
form
\begin{equation}\label{G approx}
G(h)=\int_0^1dx
h(x)\bar{\rho}+\frac{1}{2}\int_0^1dx\int_0^1dyh(x)c(x,y)h(y)+o(h^2)
\end{equation}
where $c(x,y)$ is related to $k(x,y)$ by the following relation
\begin{equation}
\int c(x,y)k(y,z) dy=\delta(x-z)
\label{inv}
\end{equation}
By inserting (\ref{G approx}) in (\ref{H_J,G}) we obtain 
the following equation for $c(x,y)$
\begin{equation}
\frac{1}{2}\partial_x^2c(x,y)- (d_1 - b_1)c(x,y
-\frac{1}{2}\bar{\rho}(1-\bar{\rho})\partial_x^2\delta(x-y)
+b_0\delta(x-y)=0
\label{ce}
\end{equation}
where $$\begin{array}{ll}
b_1=B'(\rho)_{|_{\rho=\bar{\rho}}}, &
d_1=D'(\rho)_{|_{\rho=\bar{\rho}}}\end{array}$$ and
\begin{equation}
b_0=B(\bar{\rho})=D(\bar{\rho})=d_0
\end{equation}
It is now easy to find the equation for $k(x,y)$ by combining
(\ref{ce}) with (\ref{inv}) 
\begin{equation}
\frac{1}{2}\bar{\rho}(1 - \bar{\rho}) \partial_x^2k(x,y)-
b_0k(x,y) - \frac{1}{2}\partial_x^2\delta(x-y)
+ (d_1 - b_1)\delta(x - y)=0
\label{ke}
\end{equation}
If the entropy is a local functional of the density, $k(x,y)$ has the
form  $k(x,y)=f(\bar{\rho})\delta(x-y)$ which inserted in (\ref{ke})
gives
\begin{equation}
f(\bar{\rho})=[\bar{\rho}(1-\bar{\rho})]^{-1}
\end{equation}
and
\begin{equation}
b_0[\bar{\rho}(1-\bar{\rho})]^{-1}-(d_1-b_1)=0.
\end{equation}
Therefore if $b_0,b_1,d_1$ do not satisfy this equation
the entropy cannot be a local functional of the density.

The equation for $c(x,y)$ simplifies by writing
\begin{equation}\label{c(x,y)}
c(x,y)=\bar{\rho}(1-\bar{\rho})\delta(x-y)+b(x,y)
\end{equation}
obtaining the following equation for $b(x,y)$
\begin{equation}
-\frac{1}{2}\partial_x^2b(x,y)+(d_1 - b_1)b(x,y)=
(\bar{\rho}(1-\bar{\rho})(b_1 - d_1)+b_0)\delta(x-y)
\label{cb}
\end{equation}
Remark that $d_1 - b_1$, being the second derivative of
the potential calculated in a minimum, is positive and
the solution of equation (\ref{cb}) is
exponentially decreasing. The macroscopic correlations
in the equilibrium states of our system are therefore of short range when
compared to the stationary non equilibrium correlations of the 
simple exclusion process considered in \cite{BDGJL}.

In \cite{DFL} the gaussian process
describing, in the same model, central limit type fluctuations
was studied.  When the $t \rightarrow \infty$ limit 
is taken one finds that the stationary correlations of such
a process agree, as one could expect, with
the macroscopic correlations we have calculated
in the gaussian approximation. 

\section{Conclusions}
The calculation presented in this paper supports the
conjecture that macroscopic correlations are
a generic feature of equilibrium states of non reversible 
lattice gases.
It shows in addition that the Hamilton-Jacobi equation provides 
an effective approach to the
study of the macroscopic entropy in cases which are more complex
than the conservative lattice gases considered in \cite{BDGJL}.

\acknowledgments
The continued and fruitful collaboration
on statistical mechanical problems of one of us (GJ-L) with L. Bertini, 
A. De Sole, D. Gabrielli and C. Landim  
provided the stimulating environment for this work. We are grateful to 
L. Bertini for a critical reading of the manuscript. 
We thank COFIN MIUR 2002027798 for financial support.

\end{document}